# Electric Polarization in YCrO₃ Induced by Restricted Polar Domains of Magnetic and Structural Natures

V.A. Sanina *a*,\*, B.Kh. Khannanov *a*, E.I. Golovenchits *a*, and M. P. Shcheglov *a*

*a  Ioffe Institute, St. Petersburg, 194021 Russia*

\**e-mail: sanina@mail.ioffe.ru*

**Abstract**—The electric polarization induced by local polar domains of two types (phase separation domains of magnetic nature and structure-distorted domains) has been observed in YCrO₃ single crystals. These domains form a superparaelectric state. Below some temperatures, in the frozen superparaelectric state, the pyrocurrent maxima and the hysteresis loops with remanent polarization are observed as along axis *c* so in directions [110]. The polarization exists to the temperatures depending on the orientation of electric field with respect to the crystal axes. The sources of formation of such local domains are analyzed and their properties are studied.

1. INTRODUCTION

In [1], it was reported that orthochromites RCrO₃ with magnetic rare-earth R ions are multiferroics of the second type. The polar order in them was observed below Neel temperature $T_N$ of magnetic ordering. Temperatures $T_N$ of orthochromites are quite high (130–250 K) and are substantially higher than the Curie temperatures $T_C$ of the ferroelectric ordering for the multiferroics of II-type studied before, in which the ferroelectric ordering is induced by the special type magnetic ordering [2–7]. Slightly later the polar order was detected in LuCrO₃ with a nonmagnetic R ion [8]. In this case, the polar ordering temperature $T_C$ was significantly higher than $T_N$.

RCrO₃ orthochromites have the structure of a rhombically distorted perovskite with centrosymmetrical space group *Pbnm* [9, 10] forbidding the ferroelectric ordering. The unit cell contains four formula units. $Cr^{3+}$ ions are located in the oxygen octahedra. However, the axes of the octahedra are deviated from the axis *c* along which they are oriented in undistorted cubic perovskites. Ions $R^{3+}$ are located in strongly distorted polyhedra with 8 nearest oxygen ions. The symmetry of local positions of $R^{3+}$ ($C_s$) is noncentral and, therefore, $RO_8$ quasi-molecules have electric dipole moments located in (001) planes. However, these dipole moments in the unit cell are deflected in different directions compensating each other. In this case, the antiferroelectric ordering arises that agrees with space group *Pbnm*.

The magnetic properties and the magnetic phase transitions in $RCrO_3$ were studied in detail before [9,10]. In $YCrO_3$ (YCO), at all temperatures $T \leq T_N$, the $\Gamma 4$ (*GxFz*) structure is realized (following to the Bertaut denotations [9]), i.e., there are antiferromagnetic ordering of $Cr^{3+}$ ion spins along axis $x(a)$ and the weak ferromagnetic moment *Fz* along axis $z(c)$. Figure 1 shows the temperature dependence of the magnetization *Fz* in a magnetic field $H = 5$ kOe, $H \parallel c$. It is seen that there is a well-defined magnetic phase transition near $T_N = 142$ K.

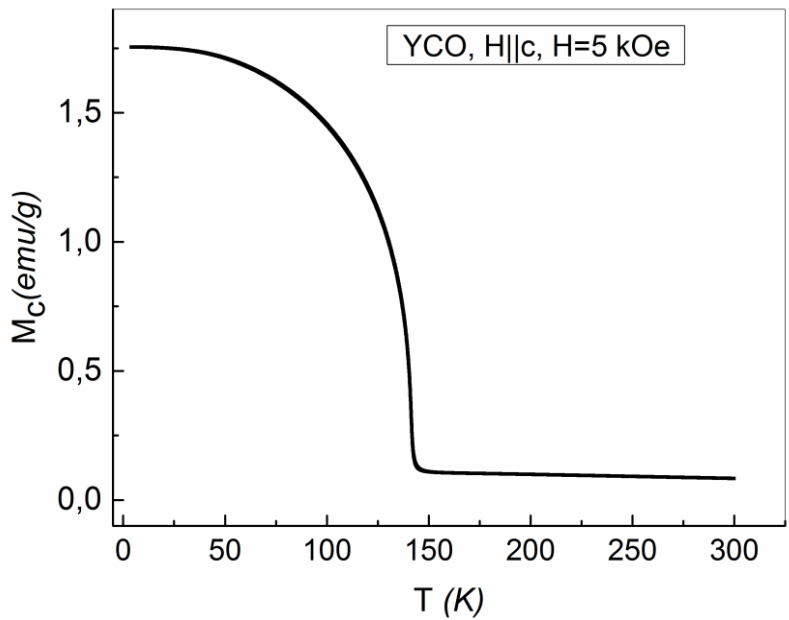

**Fig. 1.** Temperature dependence of magnetization *Fz* in YCO in magnetic field $H = 5$ kOe, $H \parallel c$.

Because it was stated in [1, 8] that the ferroelectric ordering forms in $RCrO_3$ below temperature $T_C$, then, near $T_C$, the structural phase transition to the non-centric symmetry accompanied by free dispersion maxima of the dielectric permittivity ε' and the dielectric losses ε'' must take place. However, no structural transition was observed near $T_C$ in $RCrO_3$. Note that the problems with the structural eccentricity appeared also before during studies of the low-temperature multiferroics of II-type. They had a low polarization, which was due to a small distortion of the lattice centrality that is difficult to be fixed experimentally. However, in this case, the well-defined free dispersion maxima of the dielectric permittivity were observed near $T_C$, indicating the formation of the ferroelectric ordering. In [5, 6, 11, 12], the microscopic theoretical models were developed; they described the formation of the structural eccentricity induced by the features of the magnetic structure in such multiferroics. In [1, 8], it was shown that $RCrO_3$ had a significantly larger electric polarization as compared to the multiferroics of II-type studied before, but, in this case, the dielectric permittivity anomalies near $T_C$ were significantly less intense. Note, that, in [1, 8], ceramic $RCrO_3$ samples were studied, and the verification of the obtained results on the single crystals was required.

In this work, the polar properties of $YCrO_3$ (YCO) single crystals with nonmagnetic R ion are comprehensively studied for the first time. The main objectives were a search for a possible electric polarization and establishment of its nature. Based on the studies of the dielectric permittivity, the conductivity, and the electric polarization by the pyrocurrent and hysteresis loop methods, the electric polarization in YCO was detected both along axis $c$ and in direction [110]. The polarization existed up to temperatures depending on the directions of applying electric field with respect to the crystal axes. These temperatures do not coincide with $T_N$. The observed electric polarization is found to be induced by the restricted polar domains of two types (of magnetic and structural nature), and it is not provided by homogeneous ferroelectric ordering. We analyze the sources of formation and the properties of such restricted domains.

In [13–15], the electric polarization was observed in multiferroics–manganites $RMn_2O_5$ (R = Gd, Bi) and $Gd_{0.8}Ce_{0.2}Mn_2O_5$. At temperatures $T \leq T_C \approx T_N$, the ferroelectric ordering was observed along axis $b$. In parallel with this, the hysteresis loops and maxima of pyrocurrent induced by the restricted polar phase separation domains in the frozen superparaelectric state were observed in these multiferroics over the wide temperature range 5 K $< T \leq T_{\mathrm{fr}}$ (this state was considered theoretically in [16]). The values of $T_{\mathrm{fr}}$ were significantly higher than $T_C$ and were dependent on the orientation of the crystal axes. As a result, it was demonstrated in [13–15] that the existence of the pyrocurrent maxima and the electric polarization loops are necessary but not sufficient conditions of the formation of the ferroelectric ordering.

2. EXPERIMENTAL

The YCO single crystals were grown by the method of spontaneous crystallization in a solution–melt and were characterized by symmetry *Pbnm* with lattice parameters $a = 5.243$ Å, $b = 5.524$ Å, $c = 7.536$ Å. They were 2–3-mm-thick plates with the area of 3–5 mm². The developed planes were perpendicular to either axis $c$ or direction [110]. To measure the dielectric permittivity, the conductivity, and the electric polarization, the ~0.5–1.5-mm-thick flat capacitors with the area of 3–4 mm² was used. The conductivity and the capacity were measured using a Good Will LCR-819 impedance meter in the frequency range 0.5–50 kHz in the temperature range 5–450 K. The electric polarization was measured by two methods: the pyrocurrent method and the PUND (Positive Up Negative Down) method of measuring the hysteresis loops [17]. In the first case, the polarization was measured by a Keithly 6514 electrometer during sample heating at a constant rate of varying temperature after previous cooling a sample in the polarizing electric field $E$. The PUND method was used to measure the dynamic polarization of the restricted polar domains as a response to a series two positive and two negative electric field pulses applied successively. We used the PUND method adapted for the measurement of the restricted polar regions that was used before in [13–15]. The comparative analysis of the property of the polarizations provided by the restricted polar domains measured by these two methods in $GdMn_2O_5$ and $Gd_{0.8}Ce_{0.2}Mn_2O_5$ was given in [15]. To study the properties of the restricted domains at temperatures 5–40 K, we also studied the microwave magnetic dynamics of YCO on the transmission type magnetic resonance spectrometer in the frequency range 28–40 GHz with low-frequency magnetic modulation described in [18–20]. The fine structure of the Bragg peaks was studied on the 3-crystal X-ray high-resolution diffractometer described in [21]. The magnetization was measured on a Quantum Design PPMS magnetometer. To study the influence of a magnetic field on the electric polarization, a superconducting magnet with $H \leq 8$ T was used.

## 3. EXPERIMENTAL RESULTS AND DISCUSSION

*3.1. Dielectric Permittivity and the Conductivity of YCO*

The temperature dependences of the dielectric permittivity $\varepsilon'$ in YCO along axis *c* and in transverse plane (110) for a number of frequencies are shown in Figs. 2a and 2b, respectively. The insets in the Figs. 2a and 2b show the temperature dependences for the corresponding components of conductivity $\sigma$. No free dispersion maxima of $\varepsilon'$ and $\sigma$ are observed over entire temperature range for both the crystal orientations, which demonstrates the absence of ferroelectric phase transitions in YCO up to temperatures of 350–400 K. As is seen, there are strong anisotropies of $\varepsilon'$ and $\sigma$. Along axis *c*, values of $\varepsilon'$ are four times higher and $\sigma$ are two orders of magnitude higher than these values in direction [110].

We are dealing with the real part of the conductivity $\sigma_1 = \omega\varepsilon''\varepsilon_0$ [22] (that was measured by $\tan\delta = \varepsilon''/\varepsilon'$). Here, $\omega$ is the angular frequency and $\varepsilon_0$ is $\varepsilon'$ in vacuum. Conductivity $\sigma_1$ (denoted $\sigma$) is dependent both on frequency and temperature. The low-frequency part of the conductivity is free dispersion and is referred to percolation conductivity $\sigma_{dc}$. In our case, conductivity $\sigma_{ac}$ in both directions exhibits a frequency dispersion (the insets in Figs. 2a, 2b). At temperatures $T > 180$ K, the higher the frequency, the higher is the conductivity. This frequency dispersion is characteristic of the local conductivity inside restricted domains with a barrier at their boundaries [22]. The percolation conductivity belongs to the crystal matrix. As is seen from the insets in Figs. 2a, 2b, the low-frequency conductivities (percolation conductivity) coincide in both directions ($\sim 10^{-7}(\Omega\text{ cm})^{-1}$) and are slightly changed with temperature at $T < 350$ K, which characterizes quite high dielectric characteristics of the matrix of the YCO single crystals. On the other hand, the existence of the local conductivity and the increase in $\varepsilon'$ in YCO demonstrates the existence of the restricted polar domains. To analyze the properties of the local conductivity, it is convenient to introduce the relative quantity $\sigma_{loc} = (\sigma_{ac} - \sigma_{dc})/\sigma_{dc}$ that characterizes the relationship of the local and percolation conductivities [22]. Figures 2c and 2d show the temperature dependences of $\sigma_{loc}$ at fixed frequencies along axis $c$ and in directions [110], respectively. In direction [110], $\sigma_{loc}$ appears at $T > 200$ K and exists up to temperatures higher than 400 K (Fig. 2d). There are comparatively low maxima of $\sigma_{loc}$ at temperatures that are only slightly dependent on frequency. This indicates that there is a high activation barrier at the boundaries of such polar domains (higher than 1.5 eV). At $T < 200$ K, the frequency dispersion has the opposite sign: the higher frequency the lower is the conductivity; i.e., the percolation conductivity prevails (Fig. 2d). Along axis $c$, at $T < 180$ K, the domains of the local conductivity with a low activation barrier $E_A \approx 0.2$ eV (Fig. 2c) are observed. At $T > 180$ K, the kinetic energy of charge carriers in these regions becomes sufficient to overcome such barriers. As a result, in the temperature range 180–220 K, the almost free dispersion percolation conductivity appears; in this case, the carriers, spreading over the crystal matrix at $T > 220$ K will overcome higher barriers at the boundaries of the restricted polar domains of different nature. They are the domains with the activation barrier of 0.49 eV along the same axis $c$ (Fig. 2c) and the domains with barriers of 1.5 eV in the direction perpendicular to axis $c$ (Fig. 2d). In this case, the significant increase in the local conductivity is observed along axis $c$ (Fig. 2c).

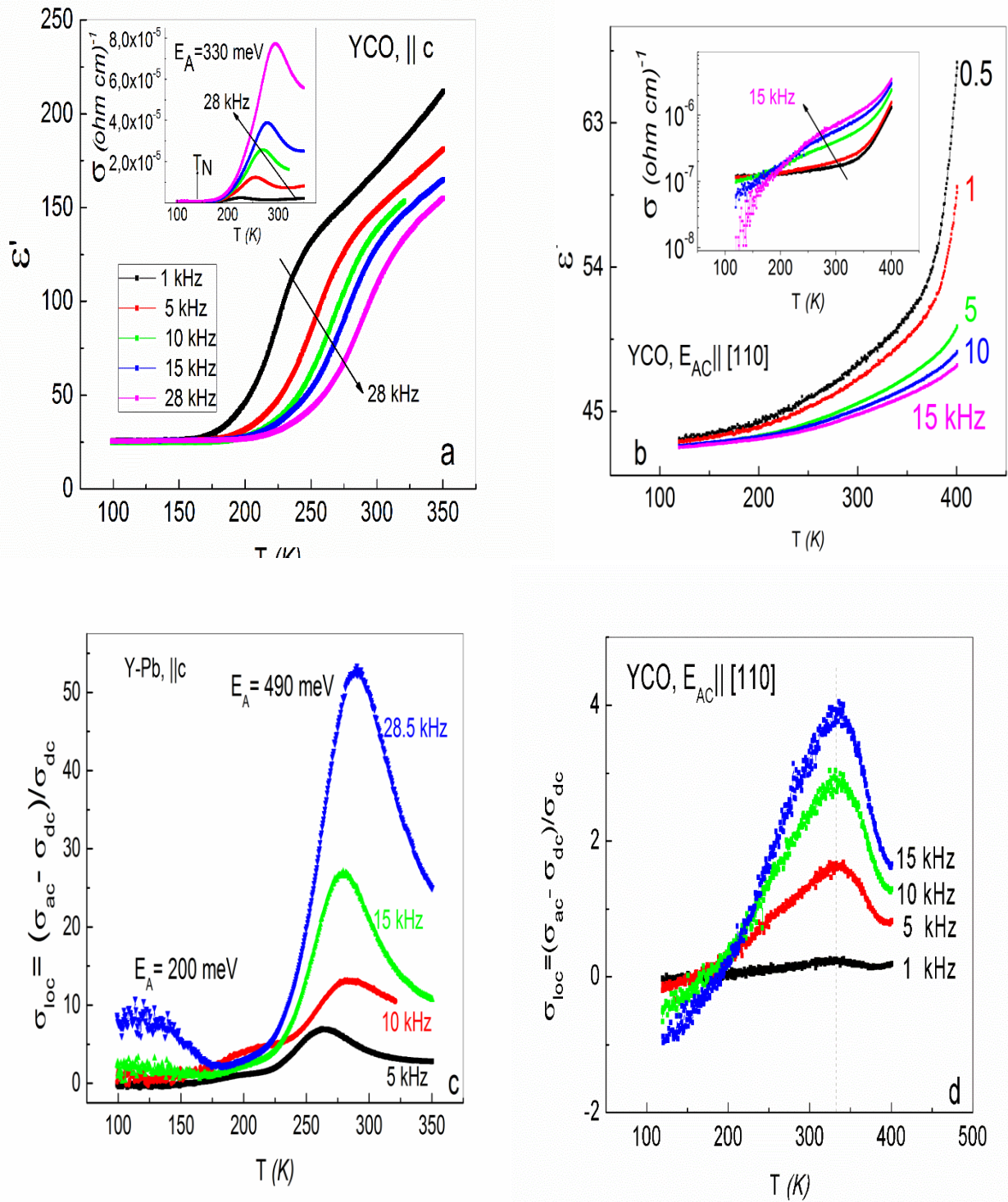

**Fig. 2.** Temperature dependences for a number of frequencies of the following values in YCO: $\varepsilon'$ (a) along axis *c*, $\varepsilon'$ (b) in the transverse plane (110); local conductivity $\sigma_{loc}$ (c) along axis *c*, $\sigma_{loc}$ (d) in the transverse plane (110). The insets to (a, b) show the temperature–frequency dependences of $\sigma$ along axis *c* and in the transverse plane (110).

We call our attention to the similarity of the properties of the restricted domains in YCO along axis $c$ at temperatures $T < 180$ K with the properties of the restricted phase separation domains in multiferroics–manganites $RMn_2O_5$ (Gd, Bi) and $Gd_{0.8}Ce_{0.2}Mn_2O_5$ containing ions of different valences $Mn^{3+}$ and $Mn^{4+}$ [13–15]. In those works, it was shown that the formation of the phase separation domains is energetically preferable and is provided by the existence of the nearest pairs of $Mn^{3+}$–$Mn4^+$ ions and finite probability of tunneling of $e_g$ electrons of $Mn^{3+}$ ions to unfilled levels of the orbital doublet of $Mn^{4+}$ ions. However, the $RMn_2O_5$ single crystals contain the same numbers of $Mn^{3+}$ and $Mn^{4+}$ ions. At the same time, the ideal YCO crystals contain only ions $Cr^{3+}$. The abovementioned similarity of the properties of the local domains suggests that the real YCO crystals also contain neighboring pairs of $Cr^{3+}$ and $Cr^{2+}$ ions that lead to the formation of the restricted phase separation domains. Note that ions $Cr^{3+}$ are analogs of ions $Mn^{4+}$ (their $3d$ shells contain three $t_{2g}$ electrons in the triplet state and an unfilled orbital doublet). On the other hand, ions $Cr^{2+}$ are analogs of ions $Mn^{3+}$ and they contain, in addition to three $t_{2g}$ electrons in the triplet state, one $e_g$ electron at the degenerate orbital doublet. It would be reasonable to assume that ions $Cr^{2+}$ appear in YCO during growing the single crystal. To grow YCO single crystals, we used the PbO–0.005PbO$_2$ solvent. The important fact is that the growth of the single crystals was carried out at a constant high temperature 1350°C of the solution–melt. In this case, in addition to initial oxides $Y_2O_3$ and $Cr_2O_3$, free ions $Pb^{2+}$ and $Pb^{4+}$ and also active oxygen ions $O^{2-}$ that recharge $Pb^{2+}$ and $Pb^{4+}$ ions emerge in the solution-melt. This leads to the change in the initial proportion of the $Pb^{2+}$ and $Pb^{4+}$ ions in the melt and the increase in the concentration of ions $Pb^{4+}$. Oxide $PbO_2$ is introduced to the solution–melt to bond active oxygen during recharging $Pb^{2+}$ and $Pb^{4+}$ ions and prevent the damage of the platinum crucible, in which the single crystals are grown. We associate the formation of the restricted polar domains in YCO with the introduction of impurity $Pb^{2+}$ and $Pb^{4+}$ ions in the position of $Y^{3+}$ ions. The ionic radii of the initial and substituting ions in positions with the eight-fold oxygen surrounding (according [23]) are as follows: 1.02 Å for $Y^{3+}$, 0.94 Å for $Pb^{4+}$, and 1.29 Å for $Pb^{2+}$. The substitution of ions $Y^{3+}$ by ions $Pb^{4+}$ leads to the appearance of ions $Cr^{2+}$ as a result of reactions $Y^{3+} = Pb^{4+} + e$, $Cr^{3+} + e = Cr^{2+}$. Thus, in YCO, in addition to $Cr^{3+}$ ions, Jahn–Teller $Cr^{2+}$ ions appear in the octahedral surrounding leading to local deformations of these octahedra. The appearance of neighboring $Cr^{3+}$ and $Cr^{2+}$ ion pairs and a finite probability of tunneling $e_g$ electrons between these ions (double exchange [24]) stimulate the energetically beneficial process of the formation of dynamically equilibrated restricted phase separation domains. These domains accumulate ferromagnetic $Cr^{3+}$–$Cr^{2+}$ ion pairs and the charge carriers (the $e_g$ electrons recharging these pairs) by analogy with that how it takes place in $LaAMnO_3$ (A = Sr, Ca, and Ba) [25, 26] and in $RMn_2O_5$ [21, 27]. The double exchange requires the existence of ferromagnetic moments $Fz$ of $Cr^{3+}$–$Cr^{2+}$ ion pair spins, which is realized at $T < T_N$ along axis $c$. In this connection, it is reasonable to refer the low-temperature regions of the local conductivity along axis $c$ to the phase separation domains that form near $Pb^{4+}$ ions (Fig. 2c). These domains with deformed oxygen octahedra violate the compensation of electric dipoles in the antiferroelectric YCO matrix, leading to the formation of the restricted polar domains of the magnetic nature. In parallel with this, there is a finite probability of incorporation of larger $Pb^{2+}$ ions with alone pairs of $6s^2$ electrons in positions of $Y^{3+}$ ions. The existence of such electron pairs is known to lead to a noncentrosymmetric distortion of the $Pb^{2+}$ ion environment [28]. As a result, the $Pb^{2+}$ impurity ions in YCO also cause the violation of compensation of the polarizations in the strongly-correlated antiferroelectric matrix and the formation of electric-dipole polarons of the structural origin. Note that ions $Pb^{4+}$ (with absenting $6s^2$ electrons) only slightly distort the structural local environment but form the phase separation domains with low barriers of 0.2 eV along axis $c$ at $T < 175$ K (Fig. 2c). We believe that the restricted structural domains forming near $Pb^{2+}$ ions are responsible for the local conductivity along axis $c$ in the domains with barriers of 0.49 eV at temperatures higher than 220 K (Fig. 2c) and the local conductivity in direction [110] in domains with the activation barrier of ~1.5 eV (Fig. 2d). Thus, in YCO, there are two types qualitatively different in their properties of the restricted polar domains that determine the properties of YCO at various orientations of the crystal and various temperatures. To determine the concentration of the impurity Pb ions, we used the X-ray fluorescence method. It turned that the Pb ion concentration in YCO was low (not higher than 1.5%), but it was sufficient to form the observed restricted polar domains.

*3.2. Electric Polarization in YCO*

Consider the measurement of the polarization by the pyrocurrent method. Figure 3a shows the temperature dependence of the electric polarization $P_c$ in YCO along axis *c*. The inset shows the corresponding dependences for the pyrocurrents. We call attention to the existence of two not split pyrocurrent maxima along axis *c* at $T < 180$ K in the first measurement cycle in field $+E \parallel c$ (the inset in Fig. 3a). During the second cycle, in the oppositely oriented field $-E \parallel c$, the alone pyrocurrent maximum is seen. In this case, the values of $P_c$ measured during these two cycles, are not completely symmetric. This situation is reproduced during the measurement on another initial sample of YCO, when, during the first cycle, after applying field $-E \parallel c$, the negative pyrocurrent maximum becomes split, and the secondary positive peaks is alone. We associate the splitting of the pyrocurrent peaks with the fact that the restricted phase separation domains of magnetic nature with small barriers forming near $Pb^{4+}$ ions are observed in YCO along the c axis at $T \leq T_N$. In the temperature range $T_N < T < 180$ K, the high-temperature slopes of the phase separation domains and the low-temperature slopes of the structural domains containing ions $Pb^{2+}$ coexist (Figs. 2c and 3a). In this case, the barriers at the boundaries of such coexisting restricted domains increase, which leads to an increase in the temperature of disappearance of the polarization as compared to $T_N$. The structural domains are longer term as compared to the phase separation magnetic domains, and they delay the decrease in the pyrocurrent. Figure 3b demonstrates the temperature dependences of the pyrocurrents in YCO in direction [110]. As is seen, the pyrocurrent maxima appeared at $T > 250$ K do not disappear up to $T \approx 375$ K, which does not allow us to correctly calculate the polarization in this direction by the pyrocurrent method. However, the non-symmetry of the values of the pyrocurrents during the measurements in two sequential cycles of applying positive and negative electric fields $E$ shows that there is the long lived polarization appearing in the first cycle that was only partially repolarized in the oppositely oriented field. This implies that the restricted polar domains of the structural nature are preferably observed in direction [110].

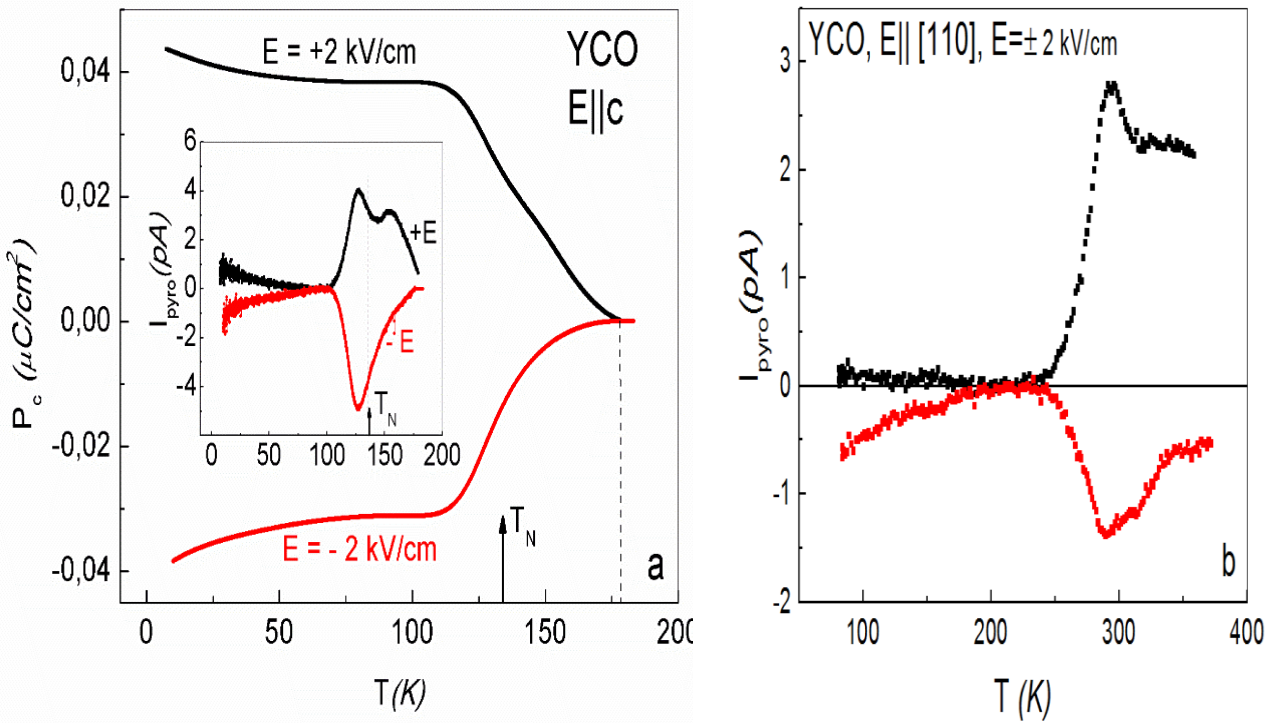

**Fig. 3.** Temperature dependences of the electric polarization and (in the inset) pyrocurrents for YCO along (a) axis $c$ and (b) pyrocurrents along direction [110].

Now we consider the measurements of the hysteresis loops of the electric polarization by the PUND method. This method, unlike the pyrocurrent method, enables one to subtract the conductivity contribution from the measured polarization. Figures 4a, 4b show the hysteresis loops of the electric polarization along axis $c$ for the most characteristic temperatures and the temperature dependence of the remanent polarization, respectively. As is seen from

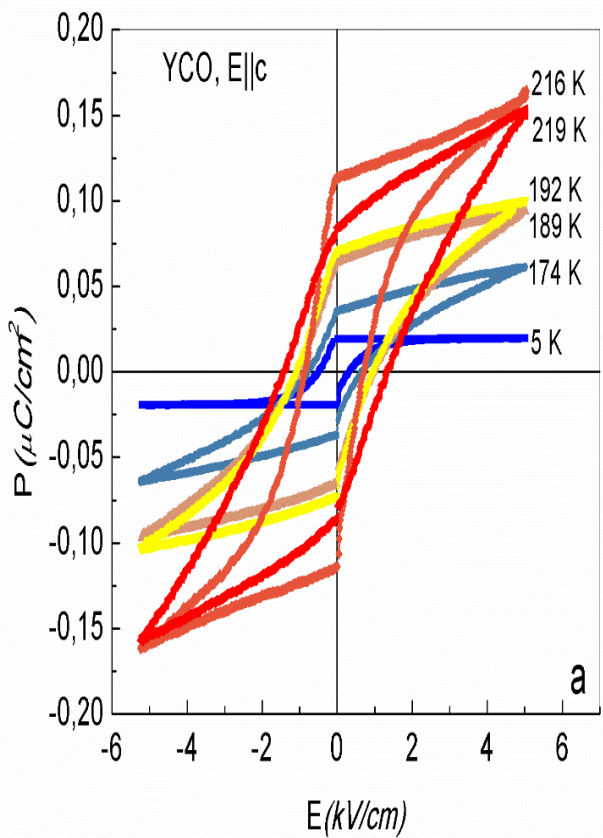
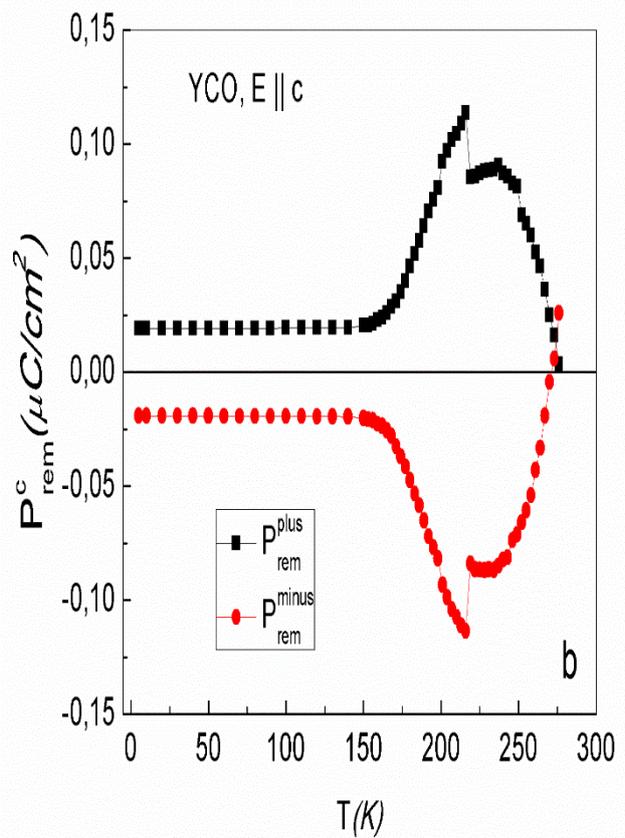
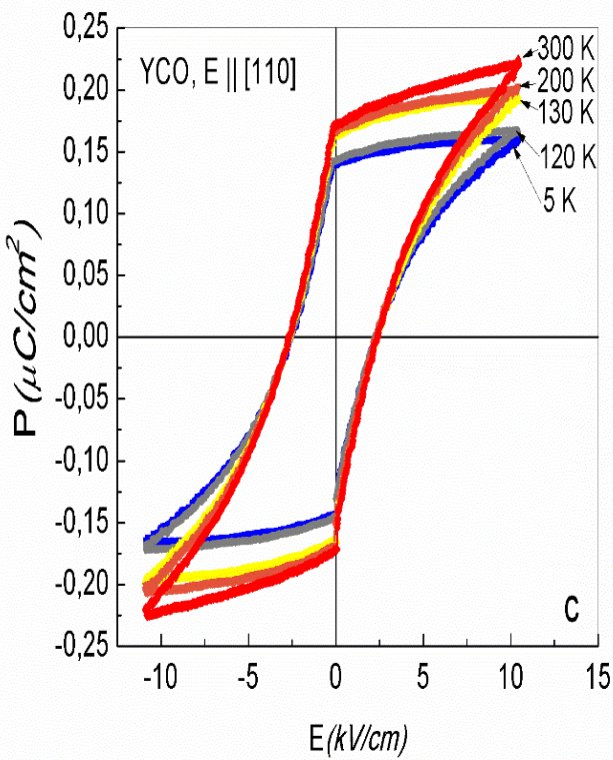
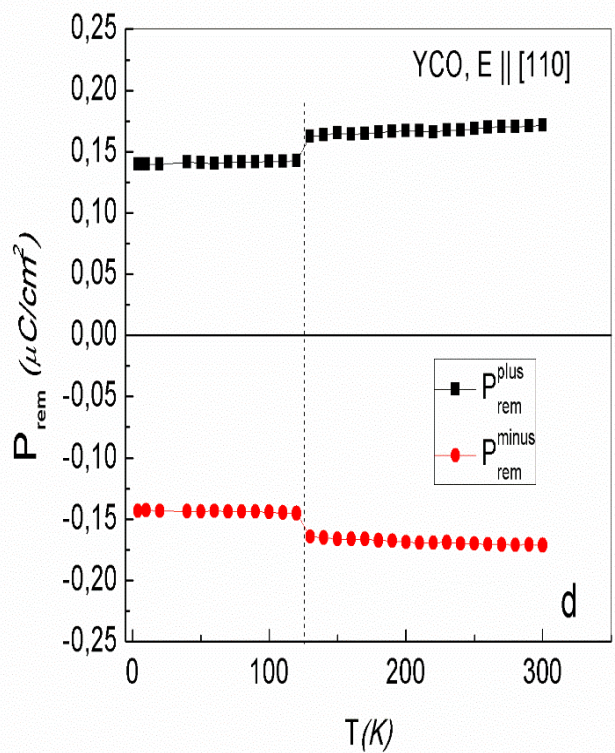

**Fig. 4.** (a) Electric polarization loops for the characteristic fixed temperatures and (b) the temperature dependence of the remanent polarization along axis *c*. Similar dependences for transverse plane (110) are shown in panels (c) and (d), respectively.

Figs. 4a, 4b, the temperature-independent saturation hysteresis loops with the remanent polarization ~0.02 μC/cm² that is two times less than that measured by the pyrotechnic method (Fig. 3a) are observed along axis $c$ in the temperature range 5– 150 K. The polarization measured by the pyrocurrent method is enhanced with the conductivity. We associate this temperature-independent polarization with the phase separation domains of the magnetic origin. As was noted when analyzing the polarization along axis $c$ measured by the pyrocurrent method, the phase separation domain and the structural domains coexist in the temperature range 175 K $> T > T_N$, which slightly increases the summary polarization. At $T > 175$ K, the local conductivity of the phase separation domains disappears, and the localized $e_g$ electrons of the phase separation domains are transformed to the percolation conductivity spreading over the crystal that contains $Pb^{4+}$ and $Pb^{2+}$ ions as before. These free electrons can be absorbed by $Pb^{4+}$ ions, transforming them to $Pb^{2+}$ ions. As a result, the concentration of the structural polar domains and the measured polarization increase. However, simultaneously, a competing process decreasing polarization $P_c$ takes place. As temperature increases, the free electron kinetic energy increases, and the probability of overcoming the barriers of 0.49 eV at the boundaries of the domains of the structural origin becomes finite, and the conductivity localized under these barriers becomes to form (Fig. 2c). It screens the internal electric field inside the structural polar domains, which decreases the polarization. The process of the transition from the increase in the polarization to the beginning of its screening occurs quite sharply in the temperature range 2–3 K (jumps in Figs. 4a, 4b). As the screening of the structural domains increases, at $T > 250$ K, the loops become more characteristic of a nonlinear dielectric and they are completely screened at 270 K (Fig. 4b).

Magnetic field $H \parallel c$ influences the magnetic phase separation domains in YCO. Field $H \parallel c$ increases moments $F_z$ of $Cr^{3+}$-$Cr^{2+}$ ion pairs and the double exchange and, thus, increases the barrier height at the boundaries of these domains, which leads to an increase in the temperature of the beginning of increasing the values of ε′ and σ (Fig. 5). Such a situation was also observed in $RMn_2O_5$ (R = Gd, Eu), leading to the increase in the polarization due to the phase separation and also to the increase in the temperature of its existence [13, 14, 21]. However, applying a magnetic field decreases the polarization in YCO (Fig. 6).

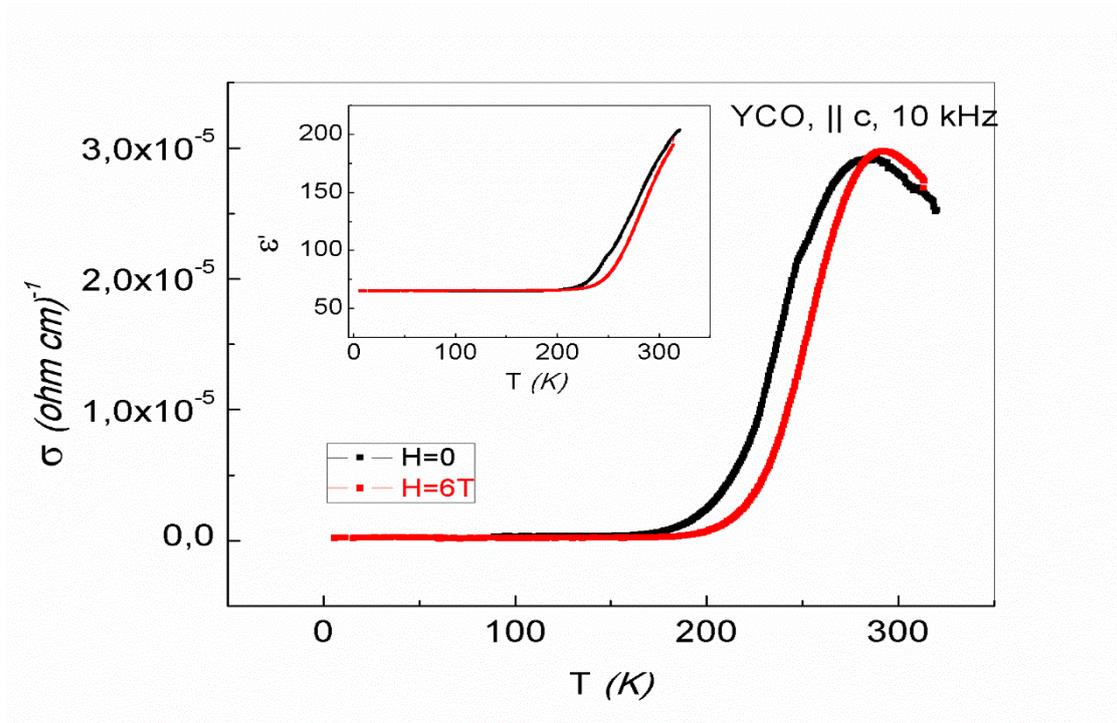

**Fig. 5.** Effect of magnetic field $H \parallel c$, $H = 6$ T on the temperature dependences of ε′ and σ for YCO

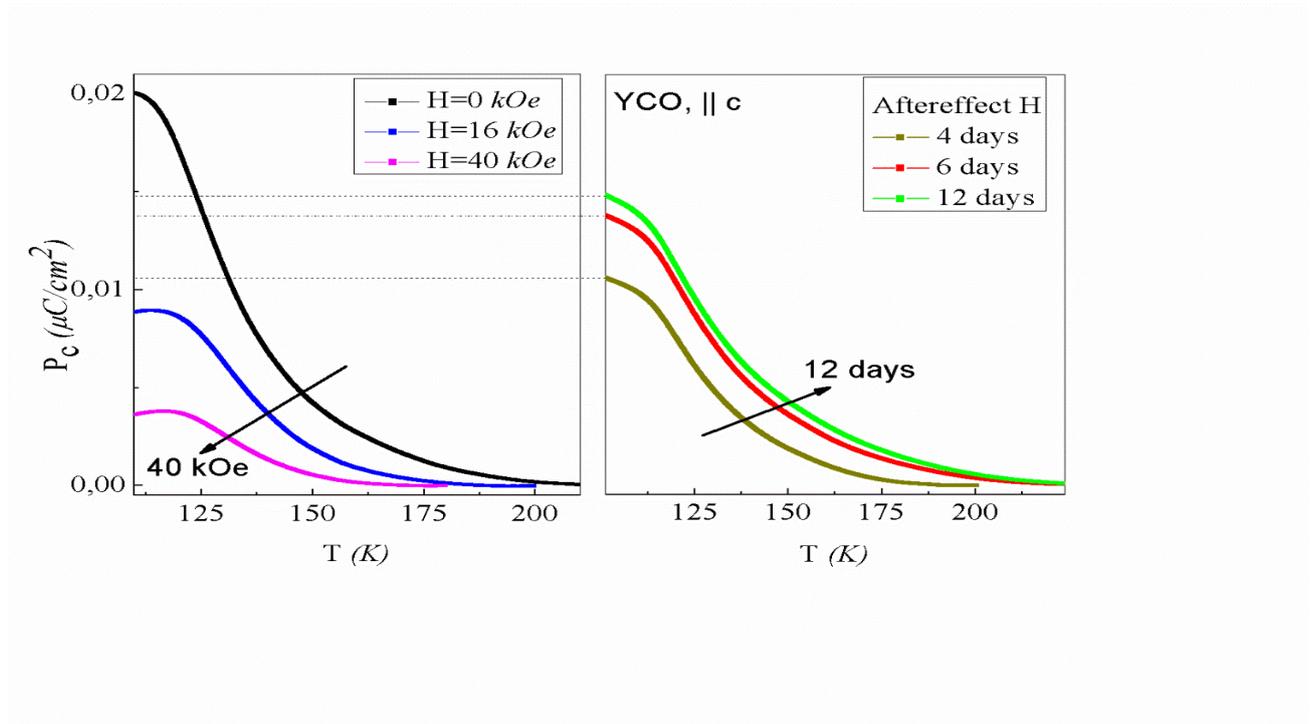

**Fig. 6.** Effect of magnetic field $H \parallel c$ on the polarization $Pc$ for YCO.

It was shown above that, in field $H = 0$ and temperature range 180 K $> T \geq T_N$, $e_g$ electrons in YCO related by the recharging process of $Cr^{3+}$–$Cr^{2+}$ ions inside the phase separation domains overcome the barriers of 0.2 eV at the boundaries of these domains, leading to an increase in the percolation conductivity of free electrons (Fig. 2c). These electrons transform $Pb^{4+}$ ions to $Pb^{2+}$ ions, accelerating damage of the polarization of the phase separation domains and, at the time, increasing the polarization of the structural domains, which leads to an increase in the summary polarization (Figs. 4a, 4b). On the other hand, the application of the magnetic field shifts the process of damaging the phase separation domains and increasing the polarization due to the structural domains. However, because the polarization of the structural domains near $Pb^{2+}$ ions is higher than that of the phase separation domains, the summary polarization decreases when applying magnetic field. Note that the process of redistribution of impurity ions $Pb^{4+}$ and $Pb^{2+}$ in magnetic fields leads to long-lived changes in the restricted polar domains and, therefore, the polarization (Fig. 6).

In the transverse direction [110], the structural restricted polar domains forming around individual $Pb^{2+}$ ions give the largest contribution to the electric polarization. In this direction, the formation of the phase separation domains at low temperatures is impossible because of the antiferromagnetic orientation of Cr ion spins (the double exchange is forbidden). However, at low temperatures $T < 130$ K, YCO has a low percolation conductivity (~$10^{-7}$ (Ohmcm)$^{-1}$) (Fig. 2d) that slightly screens the polarization (Figs. 4c, 4d). At $T > 130$ K, the polarization slightly increases with temperature and exists to temperatures higher than room temperature. Note that the polarization formed by the polar structural domains around $Pb^{2+}$ ions is one order of magnitude higher than the polarization formed by the magnetic phase separation domains that exists only along axis $c$ to a temperature that 40 K higher than $T_N$. This fact demonstrates that the violation of the compensation of the polarization in strongly-correlated antiferroelectric matrix by the structural polar domains is significantly stronger.

Note that the polarization in YCO formed predominantly by the structural domains around $Pb^{2+}$ ions (along axis $c$ at $T \approx 216$ K (Fig. 4b) and in direction [110] (Fig. 4d) are close in the value. The anisotropy of the polarization observed is really due to the existence of the phase separation formed near $Pb^{4+}$ near axis $c$.

In [18–20], it was observed in $RMn_2O_5$ that the magnetic phase separation domains at temperatures $T < 30$ K have the shape of 1D superlattices consisting of the ferromagnetic layers containing ions $Mn^{3+}$ and $Mn^{4+}$ in various proportions. In these works, we observed a characteristic set of ferromagnetic resonances from individual layers of the superlattices for all $RMn_2O_5$ independent of the type of ions R. Figure 7 shows a similar set of the ferromagnetic resonances observed in YCO that is due to the existence of the magnetic restricted phase separation domains. This confirms the fact of the existence of such domains in YCO.

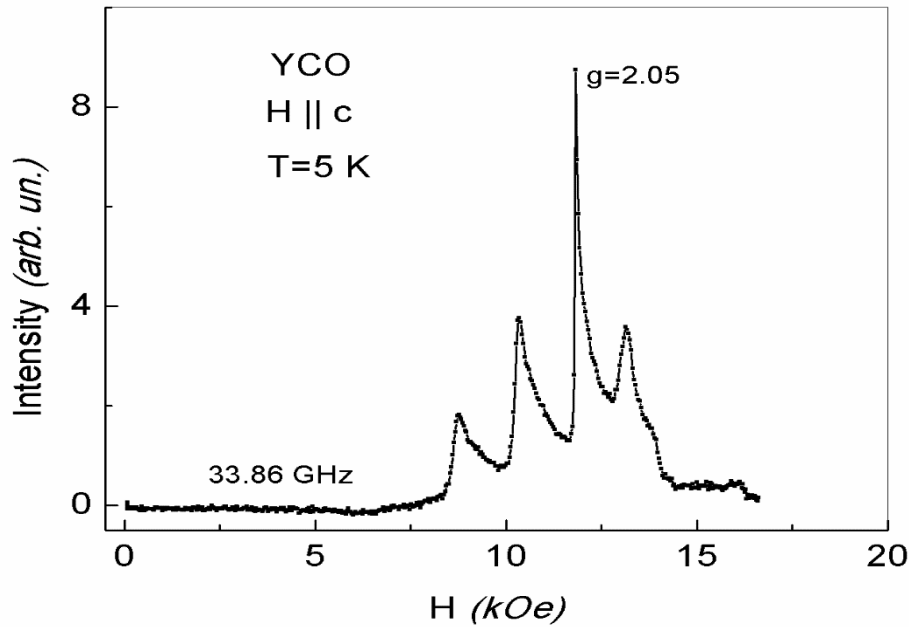

**Fig. 7.** Set of ferromagnetic resonances of the set of layers of 1D superlattices of the phase separation domain layers for YCO for axis $c$ at $T = 5$ K. The $g$-factor of the most intense ferromagnetic resonance line is 2.05.

On the other hand, the study of the fine structure of the Bragg peak (008) in YCO at room temperature on the 3-crystal high-resolution X-ray diffractometer showed the presence of well-defined additional peaks non-split with the main narrow YCO crystal matrix peak. We attribute them to well-structured local polar domains with other, but very close, inter-planar distances (Fig. 8). These domains have enough sizes and electric-dipole correlation radii for the formation of well-formed structures. Random isolated structural defects would lead to a total broadening of the main Bragg peak.

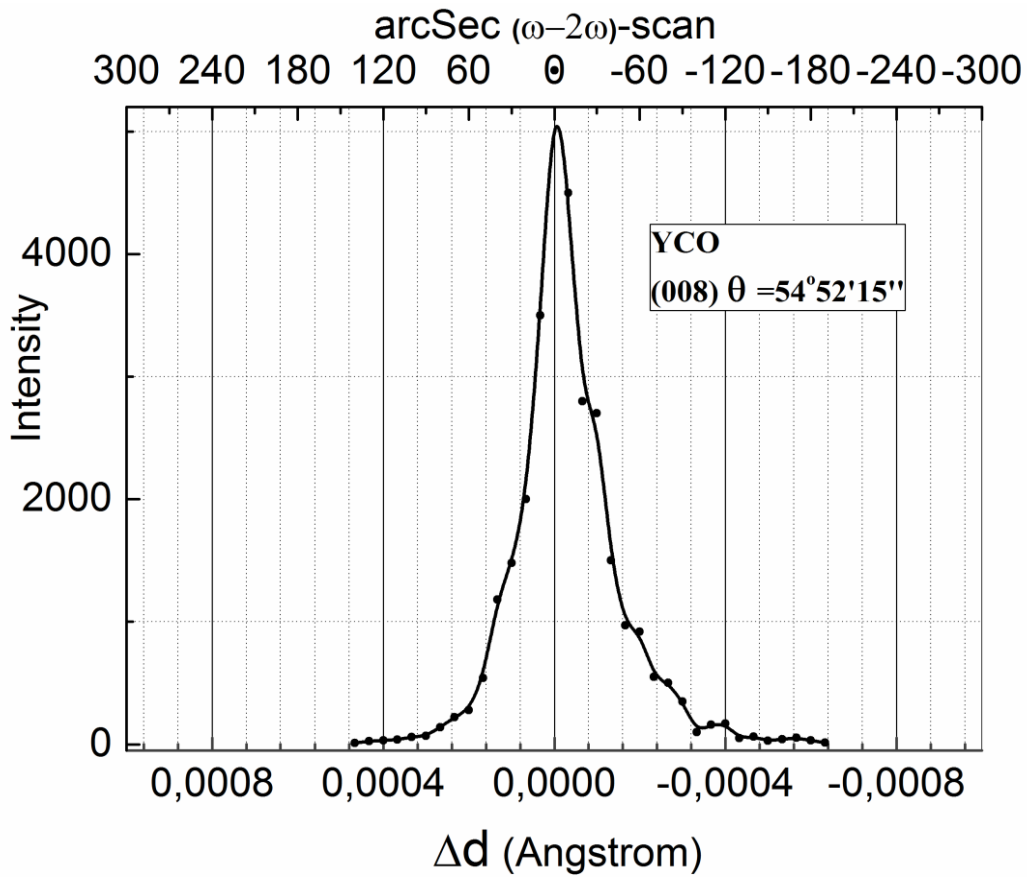

**Fig. 8.** Distribution of the Bragg reflection (008) intensities in YCO at room temperature. The lower axis shows the shift of inter-planar distances with respect to the reflection center (in angstroms). The upper axis is the crystal rotation arc in the transverse plane to the chosen direction of the reflection (in angular seconds).

We believe that the local polar domains of magnetic and also structural natures in YCO form superparaelectric state. At quite low temperatures, when $kT < E_A$ of activation barriers at the boundaries of these domains, an electric polarization (local ferroelectric state) exists and the frozen superparaelectric state, at which polarizations of these domains are not repolarized spontaneously, form in them. In this state that was considered theoretically in [16] the hysteresis loops with the remanent polarization must be observed. At temperatures $kT$fr $\approx EA$, the spontaneous repolarization of the domains appears and the common superparaelectric state appears; in this state, the remanent polarization in the hysteresis loops disappears. The analysis of the experimental results performed above shows that the conditions of formation of the frozen superparaelectric state are fulfilled for the local polar domains observed in this work. The values of temperatures $T_{fr}$ are significantly different for the magnetic and structural domains and are dependent on the direction of the crystal axes.

4. CONCLUSIONS

Thus, homogeneous ferroelectric ordering in YCO single crystals in the temperature range 5–350 K was not detected in any crystal directions. However, we observed the electric polarization induced by the local polar domains that were formed near impurity ions $Pb^{4+}$ and $Pb^{2+}$ substituting for $Y^{3+}$ ions when growing the single crystals. These impurity ions lead to the energetically beneficial processes of the formation of the local ferroelectric ordering in the isolated domains of two types in the YCO crystal matrix. They are the magnetic phase separation domains forming near $Pb^{4+}$ ions and the domains with the structural distortions forming near ions $Pb^{2+}$. The polarity of the regions is due to the violation of the compensation of the antiferroelectric state inside the domains. The polar domains form the superparaelectric state in the centrosymmetrical YCO matrix. At temperatures $T \leq T_{fr}$, at which $kT_{fr} \leq E_A$ that is the activation barrier at the polar domain boundaries, the frozen superparaelectric state forms. This state is characterized by the observation of the pyrocurrent maxima and the hysteresis loops. The electric polarization anisotropy is due to the fact that the antiferroelectric state exists in plane (001) perpendicular to axis $c$. While the phase separation domains of magnetic nature can occur only with polarization along the $c$ axis. The local polarization of the structural domains is significantly higher than similar polarization of the phase separation domains.

ACKNOWLEDGMENTS

This work was supported in part by the Russian Foundation for Basic Research (project no. 18-32-00241) and the program of the Presidium of the RAS No. 1.4 "Actual problems of the low-temperature physics."